\documentclass[fleqn,twoside]{article}
\usepackage{espcrc1}

\newcommand{\AmS}{{\protect\the\textfont2
  A\kern-.1667em\lower.5ex\hbox{M}\kern-.125emS}}

\title{Note on Evolution and Forecasting of Requirements:
  Communications Example}

\author{Mark Sh. Levin
\thanks{
 Mark Sh. Levin:~
 Inst. for Inform. Transmission Problems,
 Russian Academy of Sciences;
  http://www.mslevin.iitp.ru;
 email: mslevin@acm.org
  }
  }

\begin{document}

\maketitle

\begin{abstract}
 Combinatorial evolution and forecasting of system requirements is
 examined.
 The morphological model is used for a hierarchical requirements system
 (i.e., system parts, design alternatives for the system parts,
 ordinal estimates for the alternatives).
 A set of system changes involves changes of the
 system structure, component alternatives and their estimates.
 The composition process of the forecast is based
 on combinatorial synthesis
 (knapsack problem, multiple choice problem, hierarchical morphological design).
 An illustrative numerical example for four-phase evolution and forecasting of
 requirements to communications is described.

~~


\vspace{1pc}
\end{abstract}

\maketitle


\newcounter{cms}
\setlength{\unitlength}{1mm}

\section{Introduction}

%
%
%
 Recently, the significance of evolution and forecasting
 for communication systems has been increased (Table 1).
%



\begin{center}
 {\bf Table 1.} Some modeling problems for evolution, history evolution in communications\\
\begin{tabular}{| c |  l |l |}
\hline

 No.& Study & Source(s) \\
\hline
  1.&
 Modeling of topology evolutions and implication on proactive&\cite{wu08}\\
  &routing overhead in MANETs &\\

 2.& Historical evolution of software defined networking (SDN), its architecture
    &\cite{kreutz15,singh16}\\



  3.& Intellectual history of programmable networks (SDN)
  & \cite{feam14,nunes14}\\

 4.& History and challenges in network function virtualization
   &\cite{chow10,mij16}\\



 5.&Engineering descriptions of evolution and challenges for wireless systems
   &\cite{bhala10a,fag14,kach14,mehta14}\\

   &(\( G0 \rightarrow G1 \rightarrow  G2 \rightarrow  G3
    \rightarrow  G4 \rightarrow  G5 \rightarrow  G6\))
   &\cite{sharma13,singh07}\\

6.& Analysis of challenges and opportunities for next generation
    &\cite{hanb15,haw14}\\

 & of mobile networks&\\


\hline
\end{tabular}
\end{center}

 In general, the following three-layer framework can be examined (Fig. 1):
 (i) system requirements,
 (ii) standards,
 (iii) system(s)/product(s).
 As a result, the problems of system evolution and forecasting
 can be examined for each of the above-mentioned layers
 of communications
 (i.e., system requirements, standards, system(s)/product(s)).

\begin{center}
\begin{picture}(80,32)

\put(00,00){\makebox(0,0)[bl]{Fig. 1.
 ``Requirements-standards-systems/products''}}

\put(40,28){\oval(80,06)}


\put(015,26.5){\makebox(0,0)[bl]{Layer 1:  System requirements}}

\put(12,25){\vector(0,-1){4}} \put(22,25){\vector(0,-1){4}}
\put(32,25){\vector(0,-1){4}}

\put(54,25){\vector(0,-1){14}} \put(62,25){\vector(0,-1){14}}
\put(70,25){\vector(0,-1){14}}

\put(00,15){\line(1,0){44}} \put(00,21){\line(1,0){44}}
\put(00,15){\line(0,1){6}} \put(44,15){\line(0,1){6}}
\put(0.5,15){\line(0,1){6}} \put(43.5,15){\line(0,1){6}}

\put(07.5,16.5){\makebox(0,0)[bl]{Layer 2:  Standards}}

\put(12,15){\vector(0,-1){4}} \put(22,15){\vector(0,-1){4}}
\put(32,15){\vector(0,-1){4}}

\put(00,05){\line(1,0){80}} \put(00,11){\line(1,0){80}}
\put(00,05){\line(0,1){6}} \put(80,05){\line(0,1){6}}

\put(017,06.5){\makebox(0,0)[bl]{Layer 3: Systems/products}}

\end{picture}
\end{center}

 The article addresses combinatorial evolution and forecasting of
 requirements to communications (Table 2).
 The study of modular systems
 (systems, standards, system requirements)
 is based on morphological model (a set of part/components,
 set of design alternatives (DAs) for each part/component above,
 and ordinal estimates of DAs) \cite{lev98,lev06,lev15}.
%
%
 The system composition process is based on qualities of the selected DAs
 and qualities of their interconnections (compatibilities) (IC).
 The system composition is considered as
 combinatorial synthesis
 (knapsack problem, multiple choice problem,
 hierarchical morphological multicriteria design approach HMMD)
  \cite{lev98,lev06,lev15}.
%
%
  The author's approach to combinatorial evolution and forecasting
  has been described in
 \cite{lev98,lev02vien,lev06,levprob07,lev13,lev15}.
 The corresponding applied examples of the approach are pointed out in Table 3.
%
 %
 The presented illustrative numerical example for four-phase evolution and forecasting of
 communications requirements to network topology
 can be considered as a basis for analogical studies
 in communications and other domains.

\begin{center}
 {\bf Table 2.} Some problems over requirements \\
\begin{tabular}{| c |  l |l |}
\hline

 No.& Problems & Source(s) \\
\hline
  1.&Study, design/generation of system requirements
   &\cite{fern15,lou95,pohl10}\\




  2.& Modeling  of requirements (e.g., hierarchical modeling)
  &\cite{kuz05,levprob07}\\


  3.& Monitoring of requirements &\cite{kara16}\\

 4.&Management requirements for network function virtualization
   &\cite{bond14}\\


  5.& Modeling of evolution for requirements &\cite{kuz05,levprob07}\\

  6.& Forecasting of requirements & this paper\\

\hline
\end{tabular}
\end{center}

\begin{center}
 {\bf Table 3.} Applications of combinatorial evolution and forecasting\\
\begin{tabular}{| c |  l | c |c|l |c|}
\hline

 No.& Applied system & Evolution&Forecasting&Source(s)&Year \\
\hline

 I.&Systems/products:&&&&\\
 1.1.&Architecture of DSS COMBI-PC&Yes&None&\cite{lev93,lev98}&1993\\



 1.2.&Electronic device for signal processing&Yes&None
    &\cite{lev06,levfeld00}&2000\\



 II.&Standards:&&&&\\

  2.1.& MPEG-like standard for multimedia
     &Yes&Yes &\cite{lev15,lev09had}&2009\\

  &information transmission&&&&\\


  2.2.&ZigBee protocol for sensor networks&Yes&Yes
   & \cite{lev15,levsib10and,levand12}&2010\\



 III.&Modular educational courses:&&&&\\


 3.1.&Education course on system engineering/
     &Yes&Yes&\cite{lev13,lev15}&2013\\

    &system design&&&&\\


 IV.&System requirements:&&&&\\
 4.1.&Requirements to communications topology
      &Yes&None&\cite{kuz05,levprob07}&2005\\

 4.2.&Requirements to communications topology
       &Yes&Yes&this paper&2017\\

             &(morphological model with design alternatives)&&&   &\\



\hline
\end{tabular}
\end{center}

\section{Framework of combinatorial system evolution and forecasting}


 Knowledge representation in product design systems
 is systematically studied (e.g., \cite{chan13,hansen01}).
 Here, modular systems
 (or corresponding modular/composite alterantives/solutions)
  are examined as the following
 (i.e., system configuration)
 (e.g., \cite{lev06,lev15}):
 (a) a set of system elements (parts, components, modules),
 (b) a special structure over the system elements, e.g.,
 hierarchy, tree-like structure.
 In addition, system element alternatives
 can be considered (including estimates of the alternatives).

 Fig. 2 depicts a composite (modular) system, consisting
 of \(n\) parts/components/modules
 (\(P^{i}, i=\overline{1,n}\))
 and corresponding three design alternatives (DAs) for each
 part/component/module \(P^{i}\)
 \cite{lev98,lev06,lev15}.
%
%
  For DAs,
 the following information is considered
 (i.e., morphological system structure)
 (e.g., \cite{lev98,lev06,lev15}:
 (a) estimates of DAs
  (e.g., vector estimates, ordinal estimates,
   interval multiset estimates),
 (b) estimates of compatibility between DAs of different system components
 (e.g., ordinal estimates, interval multiset estimates).
 Generally,
 the following system change operations types can be studied and
 used
  \cite{lev98,lev02vien,lev06,levsib10,lev15}:

 {\bf I.} Operations for  DAs:~
 {\it 1.1.} change/ improvement of DA ~\(O_{1}\),
 {\it 1.2.} deletion of DA ~\(O_{2}\),
  {\it 1.3.} addition of DA ~\(O_{3}\),
 {\it 1.4.} aggregation of DAs ~\(O_{4}\).

 {\bf II.} Operations for  IC:~
  change/improvement of DAs compatibility IC ~\(O_{5}\).

 {\bf III.} Operations for subsystems (system parts, components):~
 {\it 3.1.} change/improvement of a system part ~\(O_{6}\),
 {\it 3.2.} deletion of a system part ~\(O_{7}\),
 {\it 3.3.} addition of a system part ~\(O_{8}\),
 {\it 3.4.} aggregation of system parts ~\(O_{9}\).

 {\bf IV.} Operations for the system configuration/structure (change/extension):~ (\(O_{10}\)).

 For each operation above,
  a set of attributes has to be examined
 (e.g., required resources, profit).
  Special  binary relations over the operations can be
 examined as well (e.g., compatibility, complementarity)
 \cite{lev06}.
%
%
 As a result, the improvement process can be considered  as
 selection and/or composition of
 the above-mentioned operations (items)
 while taking into account objective function(s) and resource
 constraint(s).
 This process can formulated as designing an improvement configuration
 (e.g., knapsack problem, multiple choice problem, HMMD).
 The considered scheme of system evolution and forecasting
 involves the following stages (Fig. 3)
 \cite{lev02vien,lev06,lev13,lev15}:

\begin{center}
\begin{picture}(100,51)
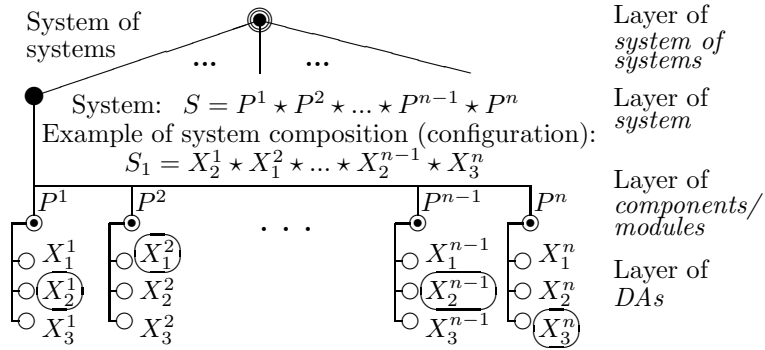
\put(00,00){\makebox(0,0)[bl] {Fig. 2. Illustration for
 modular system (morphological model) \cite{lev15}}}

\put(4,5){\makebox(0,8)[bl]{\(X^{1}_{3}\)}}
\put(4,10){\makebox(0,8)[bl]{\(X^{1}_{2}\)}}
\put(4,15){\makebox(0,8)[bl]{\(X^{1}_{1}\)}}

\put(6.4,12){\oval(6,5)}

\put(17,5){\makebox(0,8)[bl]{\(X^{2}_{3}\)}}
\put(17,10){\makebox(0,8)[bl]{\(X^{2}_{2}\)}}
\put(17,15){\makebox(0,8)[bl]{\(X^{2}_{1}\)}}

\put(19.4,17){\oval(6,5)}

\put(55,5){\makebox(0,8)[bl]{\(X^{n-1}_{3}\)}}
\put(55,10){\makebox(0,8)[bl]{\(X^{n-1}_{2}\)}}
\put(55,15){\makebox(0,8)[bl]{\(X^{n-1}_{1}\)}}

\put(59.4,12){\oval(10,5)}

\put(70,5){\makebox(0,8)[bl]{\(X^{n}_{3}\)}}
\put(70,10){\makebox(0,8)[bl]{\(X^{n}_{2}\)}}
\put(70,15){\makebox(0,8)[bl]{\(X^{n}_{1}\)}}

\put(72.4,07){\oval(6,5)}

\put(3,21){\circle*{1}} \put(16,21){\circle*{1}}
\put(54,21){\circle*{1}} \put(69,21){\circle*{1}}

\put(3,21){\circle{2}} \put(16,21){\circle{2}}
\put(54,21){\circle{2}} \put(69,21){\circle{2}}

\put(0,21){\line(1,0){02}} \put(13,21){\line(1,0){02}}
\put(51,21){\line(1,0){02}} \put(66,21){\line(1,0){02}}

\put(0,21){\line(0,-1){13}} \put(13,21){\line(0,-1){13}}
\put(51,21){\line(0,-1){13}} \put(66,21){\line(0,-1){13}}

\put(66,16){\line(1,0){01}} \put(66,12){\line(1,0){01}}
\put(66,8){\line(1,0){01}}

\put(68,16){\circle{2}}  \put(68,12){\circle{2}}
\put(68,8){\circle{2}}

\put(51,16){\line(1,0){01}} \put(51,12){\line(1,0){01}}
\put(51,8){\line(1,0){01}}

\put(53,16){\circle{2}} \put(53,12){\circle{2}}
\put(53,8){\circle{2}}

\put(13,8){\line(1,0){01}} \put(13,12){\line(1,0){01}}
\put(13,16){\line(1,0){01}}

\put(15,12){\circle{2}} \put(15,8){\circle{2}}
\put(15,16){\circle{2}}

\put(0,8){\line(1,0){01}} \put(0,12){\line(1,0){01}}
\put(0,16){\line(1,0){01}}

\put(2,12){\circle{2}} \put(2,16){\circle{2}}
\put(2,8){\circle{2}}

\put(3,26){\line(0,-1){04}} \put(16,26){\line(0,-1){04}}
\put(54,26){\line(0,-1){04}} \put(69,26){\line(0,-1){04}}

\put(3,26){\line(1,0){66}}

\put(03.4,22.6){\makebox(0,8)[bl]{\(P^{1}\) }}
\put(16.4,22.6){\makebox(0,8)[bl]{\(P^{2}\) }}

\put(33,20){\makebox(0,8)[bl]{{\bf . . .} }}

\put(54.4,22.6){\makebox(0,8)[bl]{\(P^{n-1}\) }}
\put(69.4,22.6){\makebox(0,8)[bl]{\(P^{n}\) }}

\put(02,46){\makebox(0,8)[bl]{System of }}
\put(02,42.5){\makebox(0,8)[bl]{systems }}

\put(33,48){\circle*{1.7}} \put(33,48){\circle{2.5}}
\put(33,48){\circle{3.3}}

\put(03,38){\line(3,1){30}}

\put(33,48){\line(0,-1){07}} \put(33,48){\line(4,-1){28}}

\put(24,42){\makebox(0,8)[bl]{{\bf ...} }}
\put(39,42){\makebox(0,8)[bl]{{\bf ...} }}


\put(03,26){\line(0,1){12}} \put(03,38){\circle*{2.5}}


\put(08,35){\makebox(0,8)[bl]{System:~ \(S = P^{1} \star P^{2}
 \star ... \star P^{n-1} \star P^{n} \) }}


\put(04,31){\makebox(0,8)[bl]{Example of system composition
 (configuration): }}

\put(15,27){\makebox(0,8)[bl]{\(S_{1}=X^{1}_{2}\star
 X^{2}_{1} \star ... \star X^{n-1}_{2} \star X^{n}_{3}\)}}

\put(80,47){\makebox(0,8)[bl]{Layer of}}
\put(80,43.5){\makebox(0,8)[bl]{{\it system of}}}
\put(80,41){\makebox(0,8)[bl]{{\it systems}}}

\put(80,36){\makebox(0,8)[bl]{Layer of}}
\put(80,33){\makebox(0,8)[bl]{{\it system}}}

\put(80,25){\makebox(0,8)[bl]{Layer of }}
\put(80,21.5){\makebox(0,8)[bl]{{\it components/}}}
\put(80,19.5){\makebox(0,8)[bl]{{\it modules}}}

\put(80,13){\makebox(0,8)[bl]{Layer of}}
\put(80,10){\makebox(0,8)[bl]{{\it DAs}}}

\end{picture}
\end{center}

\begin{center}
\begin{picture}(100,62)
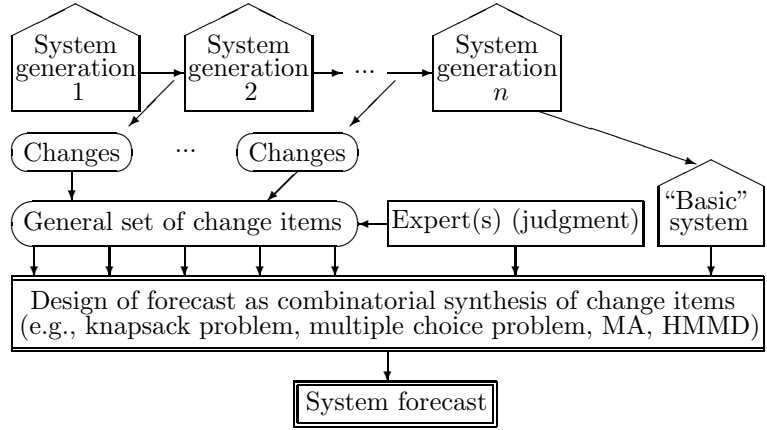
\put(00,00){\makebox(0,0)[bl]{Fig. 3. Scheme
 of evolution and forecasting (adopted from \cite{lev13,lev15}) }}


\put(00,47){\line(1,0){17}}

\put(00,47){\line(0,1){10}} \put(17,47){\line(0,1){10}}

\put(00,57){\line(2,1){08.5}} \put(17,57){\line(-2,1){08.5}}

\put(3,54){\makebox(0,0)[bl]{System}}
\put(0.5,51){\makebox(0,0)[bl]{generation}}
\put(08,48.5){\makebox(0,0)[bl]{\(1\)}}

\put(17,52){\vector(1,0){6}}


\put(23,47){\line(1,0){17}}

\put(23,47){\line(0,1){10}} \put(40,47){\line(0,1){10}}

\put(23,57){\line(2,1){08.5}} \put(40,57){\line(-2,1){08.5}}

\put(26,54){\makebox(0,0)[bl]{System}}
\put(23.5,51){\makebox(0,0)[bl]{generation}}
\put(31,48.5){\makebox(0,0)[bl]{\(2\)}}

\put(40,52){\vector(1,0){4}}

\put(45.5,52){\makebox(0,0)[bl]{...}}

\put(50,52){\vector(1,0){6}}


\put(56,47){\line(1,0){17}}

\put(56,47){\line(0,1){10}} \put(73,47){\line(0,1){10}}
\put(56,57){\line(2,1){08.5}} \put(73,57){\line(-2,1){08.5}}

\put(59,54){\makebox(0,0)[bl]{System}}
\put(56.5,51){\makebox(0,0)[bl]{generation}}
\put(64,48.5){\makebox(0,0)[bl]{\(n\)}}


\put(70,47){\vector(3,-1){21}}

\put(86,29){\line(1,0){14}}

\put(86,29){\line(0,1){08}} \put(100,29){\line(0,1){08}}

\put(86,37){\line(2,1){07}} \put(100,37){\line(-2,1){07}}

\put(86.5,34){\makebox(0,0)[bl]{``Basic''}}
\put(87.5,30.5){\makebox(0,0)[bl]{system}}

\put(93,29){\vector(0,-1){4}}

\put(50,32){\vector(-1,0){4}}

\put(50,29){\line(1,0){34}} \put(50,35){\line(1,0){34}}
\put(50,29){\line(0,1){06}} \put(84,29){\line(0,1){06}}

\put(50.5,30.5){\makebox(0,0)[bl]{Expert(s) (judgment)}}

\put(67,29){\vector(0,-1){04}}

\put(21.6,51){\vector(-1,-1){6}}

\put(08,41.5){\oval(16,5)}
\put(01.5,39.7){\makebox(0,0)[bl]{Changes}}

\put(51,51){\vector(-1,-1){6}}

\put(21.7,41.5){\makebox(0,0)[bl]{...}}

\put(38,41.5){\oval(16,5)}
\put(31.5,39.7){\makebox(0,0)[bl]{Changes}}

\put(08,39){\vector(0,-1){4}} \put(38,39){\vector(-1,-1){4}}

\put(23,32){\oval(46,6)}

\put(01.5,30.5){\makebox(0,0)[bl]{General set of change items}}

\put(23,29){\vector(0,-1){4}} \put(13,29){\vector(0,-1){4}}
\put(03,29){\vector(0,-1){4}} \put(33,29){\vector(0,-1){4}}
\put(43,29){\vector(0,-1){4}}


\put(00,15){\line(1,0){100}} \put(00,25){\line(1,0){100}}
\put(00,15.5){\line(1,0){100}} \put(00,24.5){\line(1,0){100}}
\put(00,15){\line(0,1){10}} \put(100,15){\line(0,1){10}}

\put(02.5,20){\makebox(0,0)[bl]{Design of forecast as
 combinatorial synthesis of change items}}

\put(01,16.5){\makebox(0,0)[bl]{(e.g., knapsack problem, multiple
 choice problem, MA, HMMD)}}

\put(50,15){\vector(0,-1){4}}

\put(38,05.5){\line(1,0){26}} \put(38,10.5){\line(1,0){26}}
\put(38,05.5){\line(0,1){05}} \put(64,05.5){\line(0,1){05}}


\put(39.1,06.5){\makebox(0,0)[bl]{System forecast}}

\put(37.5,05){\line(1,0){27}} \put(37.5,11){\line(1,0){27}}
\put(37.5,05){\line(0,1){06}} \put(64.5,05){\line(0,1){06}}


\end{picture}
\end{center}


 {\it Stage 1.} Analysis of chain of system generation
 and detection of system changes between neighbor system generations.

 {\it Stage 2.} Integration of system changes into a general set
 of the changes (change operations/items) while taking into account expert judgment.

 {\it Stage 3.} Selection/generation of a ``basic'' system, for example as the
 next existing system generation.

 {\it Stage 4.} Design of forecast as
 combinatorial modification of the basic system:
 system reconfiguration
 on the basis of combinatorial synthesis of change items
 (e.g., knapsack  problem, multiple choice problem,   HMMD).


\section{Example for evolution and forecasting of requirements}

 The illustrative example of
 combinatorial system evolution and forecasting
 is considered on the basis of initial data from
 \cite{kuz05,levprob07}.
 The following traditional network hierarchy can be examined:
 (a) international (multi-country, continent) network  (GAN),
 (b) metropolitan network (MN),
 (c) wide area network (WAN), and
 (d) local area network (LAN).
 Here,
 four generations for communication networks (i.e., network topological structure) are considered:

 {\it Generation 1.} Simple minimum cost network as
   one-connected structure
   (e.g., minimum cost spanning tree or minimum Steiner tree).

 {\it Generation 2.} Reliable network (e.g., bi-connected graph).

 {\it Generation 3.} Survivable network
   (e.g., bi-connected graph with additional links).

 {\it Generation 4.} Multi-layer GRID-like network environment
  (flexible, upgradeable network with reconfigurable topology).

 The
  corresponding tree-like hierarchy of requirements
 to the network (network topology) above is:


 {\bf Part 1.} User requirements \(A\):~

 {\it 1.1.} time of transmission \(T\),

 {\it 1.2.} quality (information errors, reliability of connection) \(Q\),

 {\it 1.3.} cost of transmission \(W\).

{\bf Part 2.} System requirements \(B\):

 {\it 2.1.} Basic criteria \(I\):
  2.1.1. cost \(J\),
  2.1.2. reliability \(R\),
  2.1.3. manageability \(H\),
  2.1.4. maintenanceability \(V\),
  2.1.5. testability \(E\),
  2.1.6. modularity \(M\);

 {\it 2.2.} Dynamic criteria \(Y\):
 2.2.1. adaptability \(L\),  2.2.2. safety \(F\),
 2.2.3. flexibility \(K\).

 {\bf Part 3.} Mobility requirements \(C\).

 {\bf Part 4.} Evolution/development requirements \(D\):

 {\it 4.1.} upgradeability \(U\),

 {\it 4.2.} closeness to grid \(Z\).


 For each leaf node of the structure above,
 the following design alternatives (DAs)
 (as levels of satisfiability) are examined:
 none (\(X_{0}\)),
 low level (\(X_{1}\)),
  medium level (\(X_{2}\)), and
  high level (\(X_{3}\)).
 The four-phase evolution of the hierarchical requirements structure
 is depicted in
 Fig. 4, Fig. 5, Fig. 6, and Fig. 7
 (DAs as \(X_{0}\) is absent).

\begin{center}
\begin{picture}(73,54)

\put(00,00){\makebox(0,0)[bl] {Fig. 4.
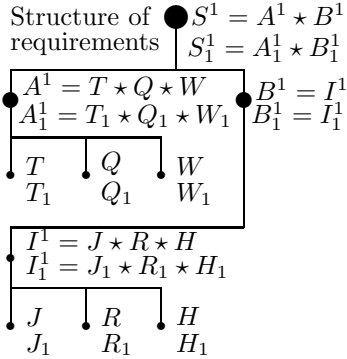
  1st system generation}}

\put(22,50){\circle*{3}}

\put(24,49){\makebox(0,0)[bl]{\(S^{1}=A^{1}\star B^{1}\)}}
\put(23.5,44){\makebox(0,0)[bl]{\(S^{1}_{1} =
 A^{1}_{1}\star B^{1}_{1}\)}}

\put(00,49){\makebox(0,0)[bl]{Structure of}}
\put(00,45){\makebox(0,0)[bl]{requirements}}

\put(22,50){\line(0,-1){7}}

\put(00,43){\line(1,0){31}}

\put(31,39){\circle*{2}}

\put(32.5,39){\makebox(0,0)[bl]{\(B^{1}=I^{1}\)}}
\put(32,35){\makebox(0,0)[bl]{\(B^{1}_{1}=I^{1}_{1}\)}}

\put(31,43){\line(0,-1){21}}

\put(00,39){\circle*{2}}

\put(00,43){\line(0,-1){14}} \put(00,34){\line(1,0){20}}

\put(0,34){\line(0,-1){5}} \put(10,34){\line(0,-1){5}}
\put(20,34){\line(0,-1){5}}

\put(00,29){\circle*{1}} \put(10,29){\circle*{1}}
\put(20,29){\circle*{1}}

\put(01.5,39){\makebox(0,0)[bl]{\(A^{1}=T\star Q\star W\)}}
\put(01,35){\makebox(0,0)[bl]{\(A^{1}_{1} =
 T_{1}\star Q_{1} \star W_{1}\)}}

\put(02,29){\makebox(0,0)[bl]{\(T\)}}
\put(02,25){\makebox(0,0)[bl]{\(T_{1}\)}}

\put(12,29){\makebox(0,0)[bl]{\(Q\)}}
\put(12,25){\makebox(0,0)[bl]{\(Q_{1}\)}}

\put(22,29){\makebox(0,0)[bl]{\(W\)}}
\put(22,25){\makebox(0,0)[bl]{\(W_{1}\)}}

\put(00,22){\line(1,0){31}}

\put(00,18){\circle*{1}}

\put(00,22){\line(0,-1){13}} \put(00,14){\line(1,0){20}}

\put(10,14){\line(0,-1){5}} \put(20,14){\line(0,-1){5}}

\put(00,9){\circle*{1}} \put(10,9){\circle*{1}}
\put(20,9){\circle*{1}}

\put(02,19){\makebox(0,0)[bl]{\(I^{1}=J\star R\star H \)}}

\put(02,15){\makebox(0,0)[bl]{\(I^{1}_{1}=J_{1}\star R_{1}
 \star H_{1}\)}}

\put(02,9){\makebox(0,0)[bl]{\(J\)}}
\put(02,05){\makebox(0,0)[bl]{\(J_{1}\)}}

\put(12,9){\makebox(0,0)[bl]{\(R\)}}
\put(12,05){\makebox(0,0)[bl]{\(R_{1}\)}}

\put(22,09){\makebox(0,0)[bl]{\(H\)}}
\put(22,05){\makebox(0,0)[bl]{\(H_{1}\)}}

\end{picture}
%
\begin{picture}(70,54)
\put(09,00){\makebox(0,0)[bl] {Fig. 5.
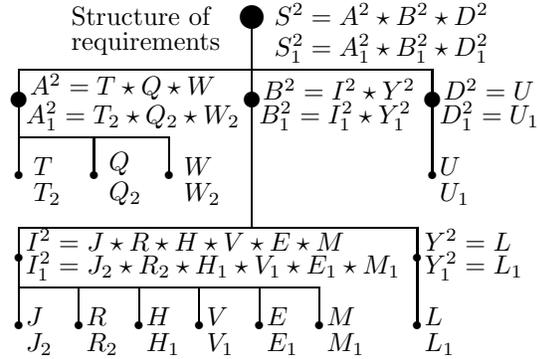
 2nd system generation}}

\put(31,50){\circle*{3}}

\put(34,49){\makebox(0,0)[bl]{\(S^{2} =
 A^{2}\star B^{2} \star D^{2}\)}}

\put(34,44){\makebox(0,0)[bl]{\(S^{2}_{1} =
  A^{2}_{1}\star B^{2}_{1}  \star D^{2}_{1}\)}}

\put(07,49){\makebox(0,0)[bl]{Structure of}}
\put(07,45){\makebox(0,0)[bl]{requirements}}

\put(31,50){\line(0,-1){7}}

\put(00,43){\line(1,0){55}}

\put(31,39){\circle*{2}}

\put(32.5,39){\makebox(0,0)[bl]{\(B^{2} =
  I^{2}\star Y^{2}\)}}

\put(32,35){\makebox(0,0)[bl]{\(B^{2}_{1} =
  I^{2}_{1}\star Y^{2}_{1}\)}}

\put(31,43){\line(0,-1){21}}


\put(55,39){\circle*{2}}

\put(55,43){\line(0,-1){14}}

\put(55,34){\line(0,-1){5}}

\put(55,29){\circle*{1}}

\put(56.5,39){\makebox(0,0)[bl]{\(D^{2} = U\)}}
\put(56,35){\makebox(0,0)[bl]{\(D^{2}_{1}=U_{1}\)}}

\put(56,29){\makebox(0,0)[bl]{\(U\)}}
\put(56,25){\makebox(0,0)[bl]{\(U_{1}\)}}


\put(00,39){\circle*{2}}

\put(00,43){\line(0,-1){14}} \put(00,34){\line(1,0){20}}

\put(0,34){\line(0,-1){5}} \put(10,34){\line(0,-1){5}}
\put(20,34){\line(0,-1){5}}

\put(00,29){\circle*{1}} \put(10,29){\circle*{1}}
\put(20,29){\circle*{1}}

\put(01.5,39){\makebox(0,0)[bl]{\(A^{2} =
 T \star Q\star W\)}}

\put(01,35){\makebox(0,0)[bl]{\(A^{2}_{1} =
 T_{2}\star Q_{2} \star W_{2}\)}}

\put(02,29){\makebox(0,0)[bl]{\(T\)}}
\put(02,25){\makebox(0,0)[bl]{\(T_{2}\)}}

\put(12,29){\makebox(0,0)[bl]{\(Q\)}}
\put(12,25){\makebox(0,0)[bl]{\(Q_{2}\)}}

\put(22,29){\makebox(0,0)[bl]{\(W\)}}
\put(22,25){\makebox(0,0)[bl]{\(W_{2}\)}}


\put(00,22){\line(1,0){53}}

\put(00,18){\circle*{1}}

\put(00,22){\line(0,-1){13}} \put(00,14){\line(1,0){40}}

\put(08,14){\line(0,-1){5}} \put(16,14){\line(0,-1){5}}
\put(24,14){\line(0,-1){5}} \put(32,14){\line(0,-1){5}}
\put(40,14){\line(0,-1){5}}

\put(00,9){\circle*{1}} \put(08,9){\circle*{1}}
\put(16,9){\circle*{1}} \put(24,9){\circle*{1}}
\put(32,9){\circle*{1}} \put(40,9){\circle*{1}}

\put(01,19){\makebox(0,0)[bl]{\(I^{2} =
 J\star R\star H\star V\star E\star M\)}}

\put(01,15){\makebox(0,0)[bl]{\(I^{2}_{1} =
 J_{2}\star R_{2}\star H_{1}\star V_{1}\star E_{1}\star M_{1}\)}}

\put(01,9){\makebox(0,0)[bl]{\(J\)}}
\put(01,05){\makebox(0,0)[bl]{\(J_{2}\)}}

\put(09,9){\makebox(0,0)[bl]{\(R\)}}
\put(09,05){\makebox(0,0)[bl]{\(R_{2}\)}}

\put(17,09){\makebox(0,0)[bl]{\(H\)}}
\put(17,05){\makebox(0,0)[bl]{\(H_{1}\)}}

\put(25,09){\makebox(0,0)[bl]{\(V\)}}
\put(25,05){\makebox(0,0)[bl]{\(V_{1}\)}}

\put(33,09){\makebox(0,0)[bl]{\(E\)}}
\put(33,05){\makebox(0,0)[bl]{\(E_{1}\)}}

\put(41,09){\makebox(0,0)[bl]{\(M\)}}
\put(41,05){\makebox(0,0)[bl]{\(M_{1}\)}}


\put(53,18){\circle*{1}}

\put(53,22){\line(0,-1){13}}

\put(53,9){\circle*{1}}

\put(54,19){\makebox(0,0)[bl]{\(Y^{2}= L\)}}

\put(54,15){\makebox(0,0)[bl]{\(Y^{2}_{1}= L_{1}\)}}

\put(54,9){\makebox(0,0)[bl]{\(L\)}}
\put(54,05){\makebox(0,0)[bl]{\(L_{1}\)}}

\end{picture}
\end{center}

\begin{center}
\begin{picture}(80,74)

\put(09,00){\makebox(0,0)[bl] {Fig. 6.
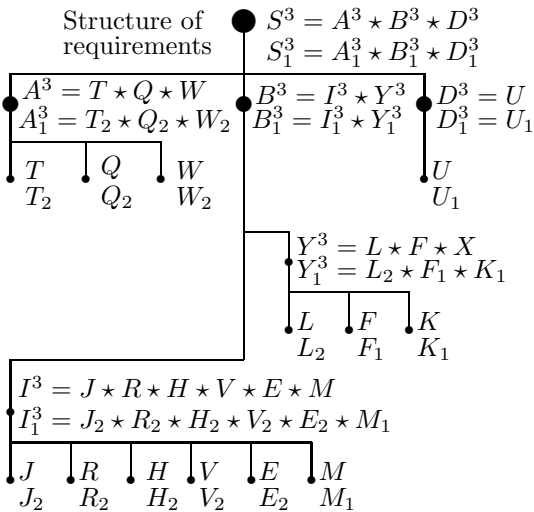
 3th system generation}}

\put(31,70){\circle*{3}}

\put(34,69){\makebox(0,0)[bl]{\(S^{3} =
  A^{3}\star B^{3}\star  D^{3}\)}}

\put(34,64){\makebox(0,0)[bl]{\(S^{3}_{1} =
 A^{3}_{1}\star B^{3}_{1} \star D^{3}_{1}\)}}

\put(07,69){\makebox(0,0)[bl]{Structure of}}
\put(07,65){\makebox(0,0)[bl]{requirements}}

\put(31,70){\line(0,-1){7}}

\put(00,63){\line(1,0){55}}

\put(31,59){\circle*{2}}

\put(32.5,59){\makebox(0,0)[bl]{\(B^{3} = I^{3}\star Y^{3}\)}}

\put(32,55){\makebox(0,0)[bl]{\(B^{3}_{1} =
  I^{3}_{1}\star Y^{3}_{1}\)}}

\put(31,63){\line(0,-1){21}}


\put(55,59){\circle*{2}}

\put(55,63){\line(0,-1){14}}

\put(55,54){\line(0,-1){5}}

\put(55,49){\circle*{1}}

\put(56.5,59){\makebox(0,0)[bl]{\(D^{3}=U\)}}
\put(56.5,55){\makebox(0,0)[bl]{\(D^{3}_{1}=U_{1}\)}}

\put(56,49){\makebox(0,0)[bl]{\(U\)}}
\put(56,45){\makebox(0,0)[bl]{\(U_{1}\)}}


\put(00,59){\circle*{2}}

\put(00,63){\line(0,-1){14}} \put(00,54){\line(1,0){20}}

\put(0,54){\line(0,-1){5}} \put(10,54){\line(0,-1){5}}
\put(20,54){\line(0,-1){5}}

\put(00,49){\circle*{1}} \put(10,49){\circle*{1}}
\put(20,49){\circle*{1}}

\put(01.5,59){\makebox(0,0)[bl]{\(A^{3} =
  T\star Q\star W\)}}

\put(01,55){\makebox(0,0)[bl]{\(A^{3}_{1} =
 T_{2}\star Q_{2} \star W_{2}\)}}

\put(02,49){\makebox(0,0)[bl]{\(T\)}}
\put(02,45){\makebox(0,0)[bl]{\(T_{2}\)}}

\put(12,49){\makebox(0,0)[bl]{\(Q\)}}
\put(12,45){\makebox(0,0)[bl]{\(Q_{2}\)}}

\put(22,49){\makebox(0,0)[bl]{\(W\)}}
\put(22,45){\makebox(0,0)[bl]{\(W_{2}\)}}


\put(00,25){\line(1,0){31}} \put(31,25){\line(0,1){17}}

\put(00,18){\circle*{1}}

\put(00,25){\line(0,-1){16}} \put(00,14){\line(1,0){40}}

\put(08,14){\line(0,-1){5}} \put(16,14){\line(0,-1){5}}
\put(24,14){\line(0,-1){5}} \put(32,14){\line(0,-1){5}}
\put(40,14){\line(0,-1){5}}

\put(00,9){\circle*{1}} \put(08,9){\circle*{1}}
\put(16,9){\circle*{1}} \put(24,9){\circle*{1}}
\put(32,9){\circle*{1}} \put(40,9){\circle*{1}}

\put(01,20){\makebox(0,0)[bl]{\(I^{3} =
 J\star R\star H\star V\star E\star M\)}}

\put(01,15){\makebox(0,0)[bl]{\(I^{3}_{1} =
 J_{2}\star R_{2}\star H_{2}\star V_{2}\star E_{2}\star M_{1}\)}}

\put(01,9){\makebox(0,0)[bl]{\(J\)}}
\put(01,05){\makebox(0,0)[bl]{\(J_{2}\)}}

\put(09,9){\makebox(0,0)[bl]{\(R\)}}
\put(09,05){\makebox(0,0)[bl]{\(R_{2}\)}}

\put(18,09){\makebox(0,0)[bl]{\(H\)}}
\put(18,05){\makebox(0,0)[bl]{\(H_{2}\)}}

\put(25,09){\makebox(0,0)[bl]{\(V\)}}
\put(25,05){\makebox(0,0)[bl]{\(V_{2}\)}}

\put(33,09){\makebox(0,0)[bl]{\(E\)}}
\put(33,05){\makebox(0,0)[bl]{\(E_{2}\)}}

\put(41,09){\makebox(0,0)[bl]{\(M\)}}
\put(41,05){\makebox(0,0)[bl]{\(M_{1}\)}}


\put(37,42){\line(-1,0){6}}

\put(37,38){\circle*{1}}

\put(37,42){\line(0,-1){13}} \put(37,34){\line(1,0){16}}
\put(45,34){\line(0,-1){5}} \put(53,34){\line(0,-1){5}}

\put(37,29){\circle*{1}} \put(45,29){\circle*{1}}
\put(53,29){\circle*{1}}

\put(38,39){\makebox(0,0)[bl]{\(Y^{3}=L\star F\star X\)}}

\put(38,35){\makebox(0,0)[bl]{\(Y^{3}_{1}=L_{2}\star F_{1}
 \star K_{1}\)}}

\put(54,29){\makebox(0,0)[bl]{\(K\)}}
\put(54,25){\makebox(0,0)[bl]{\(K_{1}\)}}

\put(46,29){\makebox(0,0)[bl]{\(F\)}}
\put(46,25){\makebox(0,0)[bl]{\(F_{1}\)}}

\put(38,29){\makebox(0,0)[bl]{\(L\)}}
\put(38,25){\makebox(0,0)[bl]{\(L_{2}\)}}

\end{picture}
%
\begin{picture}(71,80)
\put(10,00){\makebox(0,0)[bl] {Fig. 7.
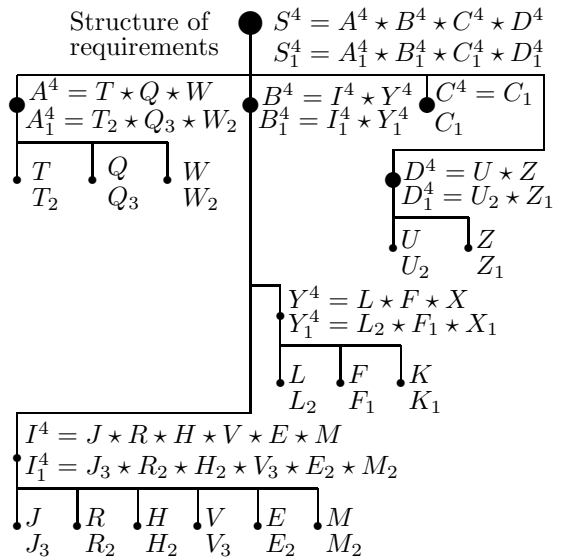
 4th system generation}}

\put(31,76){\circle*{3}}

\put(34,75){\makebox(0,0)[bl]{\(S^{4} =
  A^{4}\star B^{4}\star C^{4}\star D^{4}\)}}

\put(34,70){\makebox(0,0)[bl]{\(S^{4}_{1} =
  A^{4}_{1}\star B^{4}_{1} \star C^{4}_{1}\star D^{4}_{1}\)}}

\put(07,75){\makebox(0,0)[bl]{Structure of}}
\put(07,71){\makebox(0,0)[bl]{requirements}}

\put(31,76){\line(0,-1){7}}

\put(00,69){\line(1,0){70}}

\put(31,65){\circle*{2}}

\put(32.5,65){\makebox(0,0)[bl]{\(B^{4} = I^{4}\star Y^{4}\)}}

\put(32,61){\makebox(0,0)[bl]{\(B^{4}_{1} =
   I^{4}_{1}\star Y^{4}_{1}\)}}

\put(31,69){\line(0,-1){21}}


\put(54.5,65){\circle*{2}}

\put(55.6,65){\makebox(0,0)[bl]{\(C^{4}=C_{1}\)}}
\put(55.5,61){\makebox(0,0)[bl]{\(C_{1}\)}}

\put(54.5,69){\line(0,-1){5}}

\put(50,59){\line(1,0){20}}  \put(70,59){\line(0,1){10}}

\put(50,55){\circle*{2}}

\put(50,59){\line(0,-1){13}} \put(50,50){\line(1,0){10}}

 \put(60,50){\line(0,-1){4}}

\put(50,46){\circle*{1}} \put(60,46){\circle*{1}}

\put(51.3,55){\makebox(0,0)[bl]{\(D^{4}=U\star Z\)}}
\put(51,51){\makebox(0,0)[bl]{\(D^{4}_{1}=U_{2}\star Z_{1}\)}}

\put(51,46){\makebox(0,0)[bl]{\(U\)}}
\put(51,42){\makebox(0,0)[bl]{\(U_{2}\)}}

\put(61,46){\makebox(0,0)[bl]{\(Z\)}}
\put(61,42){\makebox(0,0)[bl]{\(Z_{1}\)}}


\put(00,65){\circle*{2}}

\put(00,69){\line(0,-1){14}} \put(00,60){\line(1,0){20}}

\put(0,60){\line(0,-1){5}} \put(10,60){\line(0,-1){5}}
\put(20,60){\line(0,-1){5}}

\put(00,55){\circle*{1}} \put(10,55){\circle*{1}}
\put(20,55){\circle*{1}}

\put(01.5,65){\makebox(0,0)[bl]{\(A^{4} =
  T\star Q\star W\)}}

\put(01,61){\makebox(0,0)[bl]{\(A^{4}_{1} =
 T_{2}\star Q_{3} \star W_{2}\)}}

\put(02,55){\makebox(0,0)[bl]{\(T\)}}
\put(02,51){\makebox(0,0)[bl]{\(T_{2}\)}}

\put(12,55){\makebox(0,0)[bl]{\(Q\)}}
\put(12,51){\makebox(0,0)[bl]{\(Q_{3}\)}}

\put(22,55){\makebox(0,0)[bl]{\(W\)}}
\put(22,51){\makebox(0,0)[bl]{\(W_{2}\)}}


\put(00,24){\line(1,0){31}} \put(31,24){\line(0,1){24}}

\put(00,18){\circle*{1}}

\put(00,24){\line(0,-1){15}} \put(00,14){\line(1,0){40}}

\put(08,14){\line(0,-1){5}} \put(16,14){\line(0,-1){5}}
\put(24,14){\line(0,-1){5}} \put(32,14){\line(0,-1){5}}
\put(40,14){\line(0,-1){5}}

\put(00,9){\circle*{1}} \put(08,9){\circle*{1}}
\put(16,9){\circle*{1}} \put(24,9){\circle*{1}}
\put(32,9){\circle*{1}} \put(40,9){\circle*{1}}

\put(01,20){\makebox(0,0)[bl]{\(I^{4} =
 J\star R\star H\star V\star E\star M\)}}

\put(01,15){\makebox(0,0)[bl]{\(I^{4}_{1} =
  J_{3}\star R_{2}\star H_{2}\star V_{3}\star E_{2}\star M_{2}\)}}

\put(01,9){\makebox(0,0)[bl]{\(J\)}}
\put(01,05){\makebox(0,0)[bl]{\(J_{3}\)}}

\put(09,9){\makebox(0,0)[bl]{\(R\)}}
\put(09,05){\makebox(0,0)[bl]{\(R_{2}\)}}

\put(17,09){\makebox(0,0)[bl]{\(H\)}}
\put(17,05){\makebox(0,0)[bl]{\(H_{2}\)}}

\put(25,09){\makebox(0,0)[bl]{\(V\)}}
\put(25,05){\makebox(0,0)[bl]{\(V_{3}\)}}

\put(33,09){\makebox(0,0)[bl]{\(E\)}}
\put(33,05){\makebox(0,0)[bl]{\(E_{2}\)}}

\put(41,09){\makebox(0,0)[bl]{\(M\)}}
\put(41,05){\makebox(0,0)[bl]{\(M_{2}\)}}

\put(35,41){\line(-1,0){4}}

\put(35,37){\circle*{1}}

\put(35,41){\line(0,-1){13}} \put(35,33){\line(1,0){16}}
\put(43,33){\line(0,-1){5}} \put(51,33){\line(0,-1){5}}

\put(35,28){\circle*{1}} \put(43,28){\circle*{1}}
\put(51,28){\circle*{1}}

\put(36,38){\makebox(0,0)[bl]{\(Y^{4} =
  L\star F\star X\)}}

\put(36,34){\makebox(0,0)[bl]{\(Y^{4}_{1} =
  L_{2}\star F_{1} \star X_{1}\)}}

\put(52,28){\makebox(0,0)[bl]{\(K\)}}
\put(52,24){\makebox(0,0)[bl]{\(K_{1}\)}}

\put(44,28){\makebox(0,0)[bl]{\(F\)}}
\put(44,24){\makebox(0,0)[bl]{\(F_{1}\)}}

\put(36,28){\makebox(0,0)[bl]{\(L\)}}
\put(36,24){\makebox(0,0)[bl]{\(L_{2}\)}}

\end{picture}
\end{center}

  The modular presentations of requirements are:

 Generation 1:
 \(S^{1}_{1} = A^{1}_{1} \star B^{1}_{1} \star C^{1}_{0} \star D^{1}_{0}  =  \)

 \(( T_{1} \star Q_{1} \star W_{1}) \star \)
 \((J_{1}\star R_{1}\star H_{1}\star V_{0}\star E_{0} \star  M_{0}) \star\)
 \((L_{0}\star F_{0}\star K_{0})\star C_{0}\star (U_{0}\star Z_{0})\).

 Generation 2:
 \(S^{2}_{1} = A^{2}_{1} \star B^{2}_{1} \star C^{2}_{0} \star D^{2}_{1} =  \)

 \(( T_{2} \star Q_{2} \star W_{2}) \star \)
 \((J_{3}\star R_{2}\star H_{1}\star V_{1}\star E_{1} \star  M_{1}) \star\)
 \((L_{1}\star F_{0}\star K_{0})\star C_{0}\star (U_{1}\star Z_{0})\).

 Generation 3:
 \(S^{3}_{1} = A^{3}_{1} \star B^{3}_{1} \star D^{3}_{1} =  \)

 \(( T_{2} \star Q_{2} \star W_{2}) \star \)
 \((J_{3}\star R_{2}\star H_{2}\star V_{3}\star E_{2} \star M_{1}) \star\)
 \((L_{2}\star F_{1}\star K_{1})\star C_{0}\star (U_{1}\star Z_{0})\).

 Generation 4:
\(S^{4}_{1} = A^{4}_{1} \star B^{4}_{1} \star C^{4}_{1} \star
 D^{4}_{1} =  \)

 \(( T_{2} \star Q_{3} \star W_{2}) \star \)
 \((J_{3}\star R_{2}\star H_{2}\star V_{3}\star E_{2} \star M_{2}) \star\)
 \((L_{2}\star F_{1}\star K_{1})\star C_{1}\star (U_{2}\star Z_{1})\).

 The local changes of the requirements
 with estimates upon two criteria
 (cost of change, \(0\) is the best value;
 profit of change, the maximum value is the best one;
 expert judgment)
 are presented in Table 4, Table 5, and Table 6.

 Now \(S^{4}\) is considered as a basis for the forecasting.
 A set of prospective change operations/items (improvements, DAs) is contained
 in Table 7
 (ordinal priorities of DAs are based on the use of multicriteria ranking,
 priority ordinal scale is \([1,3]\)).
 HMMD is used
 \cite{lev98,lev06,lev15}.
 The best composition
 of the change operations (improvements)
 is searched for as the forecast of the system requirement.
 The hierarchical structure of the composite system improvement is depicted
 in Fig. 8.
 Table 8 and Table 9 contain ordinal estimates
 of compatibilities between DAs
 (ordinal scale \([0,3]\)).
 Finally,
 two Pareto-efficient composite improvements are
 (it is assumed composite DAs for \(A^{I}\), \(B^{I}\), \(D^{I}\)
 are compatible):~

 (i) \(S^{I}_{1} = A_{1}
 \star (\widetilde{B}_{2} \star \widehat{B}_{2} ) \star
 (\widetilde{D}_{2}\star \widehat{D}_{3}\star \overline{D}_{3})\),
 here \(N(B^{I}_{1})=(3;2,0)\), \(N(D^{I}_{1})=(3;2,1,0)\));

 (ii) \(S^{I}_{2} = A_{1}
 \star (\widetilde{B}_{2} \star \widehat{B}_{2} ) \star
 (\widetilde{D}_{3}\star \widehat{D}_{3}\star \overline{D}_{3})\),
 here \(N(B^{I}_{1})=(3;2,0)\), \(N(D^{I}_{2})=(2;3,0,0)\)).

\begin{center}
 {\bf Table 4.} Change operations for \(S^{1} \Rightarrow  S^{2}\)   \\
\begin{tabular}{| c | c | c | c|}
\hline
 No.  & Change operation & Cost & Profit\\
\hline

 1.& \(J_{1} \rightarrow J_{2}\) &\(1.5\) &\(1.5\)\\

 2.& \(R_{1} \rightarrow R_{2}\) &\(2.0\) &\(2.5\)\\

 3.& \(T_{1} \rightarrow T_{2}\) &\(2.2\) &\(3.0\)\\

 4.& \(Q_{1} \rightarrow Q_{2}\) &\(1.6\) &\(2.0\)\\

 5.& \(W_{1} \rightarrow W_{2}\) &\(1.5\) &\(1.4\)\\

 6.& \(V_{0} \rightarrow V_{1}\) &\(2.0\) &\(2.1\)\\

 7.& \(E_{0} \rightarrow E_{1}\) &\(1.4\) &\(1.7\)\\

 8.& \(M_{0} \rightarrow M_{1}\) &\(1.9\) &\(1.5\)\\

 9.& \(L_{0} \rightarrow L_{1}\) &\(1.8\) &\(1.5\)\\

 10.& \(U_{0} \rightarrow U_{1}\) &\(2.0\) &\(1.6\)\\

\hline
\end{tabular}
\end{center}

 The corresponding two resultant system requirements forecasts are:

 (i) \(S^{F}_{1} = A^{F}_{1} \star B^{F}_{1} \star C^{F}_{1} \star
 D^{F}_{1} =  \)

 \(( T_{2} \star Q_{3} \star W_{2})\) \(\star \)
 \((J_{3}\star R_{2}\star H_{2}\star V_{3}\star E_{3} \star M_{3})\)\
 \( \star\)
 \((L_{2}\star F_{2}\star K_{3})\star C_{1}\star (U_{2}\star Z_{3})\);

 (ii) \(S^{F}_{1} = A^{F}_{1} \star B^{F}_{1} \star C^{F}_{1} \star
 D^{F}_{1} =  \)

 \(( T_{2} \star Q_{3} \star W_{2})\)   \( \star \)
 \((J_{3}\star R_{2}\star H_{2}\star V_{3}\star E_{3} \star M_{3})\)
 \( \star\)
 \((L_{2}\star F_{3}\star K_{3})\star C_{1}\star (U_{2}\star Z_{3})\).

\begin{center}
 {\bf Table 5.} Change operations for \(S^{2} \Rightarrow  S^{3}\)   \\
\begin{tabular}{| c | c | c | c|}
\hline
 No.  & Change operation & Cost & Profit\\
\hline
 1.& \(H_{1} \rightarrow H_{2}\) &\(2.5\) &\(2.5\)\\

 2.& \(V_{1} \rightarrow V_{2}\) &\(2.2\) &\(2.4\)\\

 3.& \(E_{1} \rightarrow E_{2}\) &\(1.5\) &\(2.0\)\\


 4.& \(L_{1} \rightarrow L_{2}\) &\(1.5\) &\(1.8\)\\

 5.& \(F_{0} \rightarrow F_{1}\) &\(1.4\) &\(2.0\)\\

 6.& \(K_{0} \rightarrow K_{1}\) &\(1.5\) &\(2.1\)\\

\hline
\end{tabular}
\end{center}

\newpage
\begin{center}
 {\bf Table 6.} Change operations for \(S^{3} \Rightarrow  S^{4}\)   \\
\begin{tabular}{| c | c | c | c|}
\hline
 No.  & Change operation & Cost & Profit\\
\hline

 1.& \(Q_{2} \rightarrow Q_{3}\) &\(2.1\) &\(3.0\)\\

 2.& \(J_{2} \rightarrow J_{3}\) &\(1.7\) &\(2.0\)\\

 3.& \(M_{1} \rightarrow M_{2}\) &\(1.6\) &\(1.8\)\\

 4.& \(U_{1} \rightarrow U_{2}\) &\(1.8\) &\(2.0\)\\

 5.& \(C_{0} \rightarrow C_{1}\) &\(2.8\) &\(3.0\)\\

 6.& \(Z_{0} \rightarrow Z_{1}\) &\(1.5\) &\(2.0\)\\

\hline
\end{tabular}
\end{center}

\begin{center}
 {\bf Table 7.} Prospective improvement for \(S^{4}\)   \\
\begin{tabular}{| c | l | c | c|c|}
\hline
 No.  & Change operation/item& Cost & Profit&Priority\\

%
\hline

 I.& Part A:&&&\\

 1.1.& \(A_{1}\): none &\(0\) &\(0\) & \(2\) \\

 1.2.& \(A_{2}\): \(W_{2} \rightarrow W_{3}\) &\(1.4\) &\(1.6\) &\(1\)\\

\hline
 II.&Part B:&&&\\

 2.1.1.& \(\widetilde{B}_{1}\): none &\(0\) &\(0\)  &\(3\)\\

 2.1.2.& \(\widetilde{B}_{2}\):  \(E_{2} \rightarrow E_{3}\) &\(2.0\) &\(2.1\) &\(1\)\\

 2.2.1.&  \(\widehat{B}_{1}\): none &\(0\) &\(0\) & \(3\) \\
 2.2.2.&  \(\widehat{B}_{2}\): \(M_{2} \rightarrow M_{3}\) &\(1.6\) &\(1.9\) &\(1\)\\

\hline
 III.& Part D:&&&\\

 3.1.1.&\(\widetilde{D}_{1}\):  none  &\(0\) &\(0\) & \(3\) \\
 3.1.2.& \(\widetilde{D}_{2}\):  \(F_{1} \rightarrow F_{2}\) &\(1.7\) &\(2.1\) & \(2\)\\
 3.1.3.& \(\widetilde{D}_{3}\): \(F_{1} \rightarrow F_{3}\) &\(2.1\) &\(3.9\)&\(1\)\\

 3.2.1.&\(\widehat{D}_{1}\):  none  &\(0\) &\(0\) &\(3\)\\

 3.2.2.& \(\widehat{D}_{2}\):  \(K_{1} \rightarrow K_{2}\) &\(1.5\) &\(2.0\)&\(2\)\\

 3.2.3.& \(\widehat{D}_{3}\): \(K_{1} \rightarrow K_{3}\) &\(3.0\) &\(4.1\)&\(1\)\\

 3.3.1.&\(\overline{D}_{1}\):  none  &\(0\) &\(0\)& \(3\) \\

 3.3.2.& \(\overline{D}_{2}\):  \(Z_{1} \rightarrow Z_{2}\) &\(1.6\) &\(2.0\)&\(2\)\\

 3.3.3.& \(\overline{D}_{3}\): \(Z_{1} \rightarrow Z_{3}\) &\(2.1\) &\(4.1\)&\(1\)\\

\hline
\end{tabular}
\end{center}

\begin{center}
\begin{picture}(80,51)

\put(00,00){\makebox(0,0)[bl] {Fig. 8.
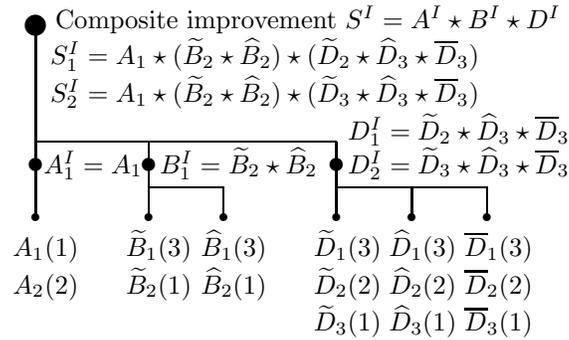
 Resultant structure of system improvement}}

\put(40,04){\makebox(0,8)[bl]{\(\widetilde{D}_{3}(1)\)}}
\put(40,09){\makebox(0,8)[bl]{\(\widetilde{D}_{2}(2)\)}}
\put(40,14){\makebox(0,8)[bl]{\(\widetilde{D}_{1}(3)\)}}

\put(50,04){\makebox(0,8)[bl]{\(\widehat{D}_{3}(1)\)}}
\put(50,09){\makebox(0,8)[bl]{\(\widehat{D}_{2}(2)\)}}
\put(50,14){\makebox(0,8)[bl]{\(\widehat{D}_{1}(3)\)}}

\put(60,04){\makebox(0,8)[bl]{\(\overline{D}_{3}(1)\)}}
\put(60,09){\makebox(0,8)[bl]{\(\overline{D}_{2}(2)\)}}
\put(60,14){\makebox(0,8)[bl]{\(\overline{D}_{1}(3)\)}}

\put(43,20){\circle*{1.2}} \put(53,20){\circle*{1.2}}
\put(63,20){\circle*{1.2}}

\put(43,20){\line(0,1){04}} \put(53,20){\line(0,1){04}}
\put(63,20){\line(0,1){04}}

\put(43,24){\line(1,0){20}}

\put(43,24){\line(0,1){06}}

\put(43,27){\circle*{1.6}}

\put(44.5,29.5){\makebox(0,8)[bl]{\(D^{I}_{1} =
 \widetilde{D}_{2}\star \widehat{D}_{3}\star \overline{D}_{3}\)}}

\put(44.5,25){\makebox(0,8)[bl]{\(D^{I}_{2} =
 \widetilde{D}_{3}\star \widehat{D}_{3}\star \overline{D}_{3}\)}}


\put(03,24){\line(0,1){06}}

\put(03,27){\circle*{1.6}}

\put(04.1,25){\makebox(0,8)[bl]{\(A^{I}_{1} = A_{1}\)}}

\put(03,20){\line(0,1){04}}

\put(03,20){\circle*{1.2}}

\put(00,14){\makebox(0,8)[bl]{\(A_{1}(1)\)}}
\put(00,09){\makebox(0,8)[bl]{\(A_{2}(2)\)}}


\put(18,24){\line(0,1){06}}

\put(18,27){\circle*{1.6}}

\put(19.5,25){\makebox(0,8)[bl]{\(B^{I}_{1} =
 \widetilde{B}_{2} \star \widehat{B}_{2} \)}}

\put(18,24){\line(1,0){10}}

\put(18,20){\line(0,1){04}} \put(28,20){\line(0,1){04}}

\put(18,20){\circle*{1.2}} \put(28,20){\circle*{1.2}}

\put(15,14){\makebox(0,8)[bl]{\(\widetilde{B}_{1}(3)\)}}
\put(15,09){\makebox(0,8)[bl]{\(\widetilde{B}_{2}(1)\)}}

\put(25,14){\makebox(0,8)[bl]{\(\widehat{B}_{1}(3)\)}}
\put(25,09){\makebox(0,8)[bl]{\(\widehat{B}_{2}(1)\)}}

\put(03,30){\line(1,0){40}} \put(03,30){\line(0,1){15.5}}

\put(03,45.5){\circle*{2.8}}

\put(05,34.5){\makebox(0,8)[bl]{\(S^{I}_{2} = A_{1}
 \star (\widetilde{B}_{2} \star \widehat{B}_{2} ) \star
 (\widetilde{D}_{3} \star \widehat{D}_{3} \star \overline{D}_{3} )\)}}

\put(05,39.5){\makebox(0,8)[bl]{\(S^{I}_{1} = A_{1}
 \star (\widetilde{B}_{2} \star \widehat{B}_{2} ) \star
 (\widetilde{D}_{2} \star \widehat{D}_{3} \star \overline{D}_{3} )\)}}

\put(05.6,44.5){\makebox(0,8)[bl]{Composite improvement
 \( S^{I} = A^{I} \star B^{I} \star D^{I} \)}}

\end{picture}
\end{center}

\newpage
\begin{center}
 {\bf Table 8.} Ordinal compatibility\\
\begin{tabular}{| l | c   c  |}
\hline
 &
 \(\widehat{B}_{1}\) & \(\widehat{B}_{2}\)  \\
\hline

 \(\widetilde{B}_{1}\) &\(2\)&\(2\)\\

 \(\widetilde{B}_{2}\) &\(2\)&\(3\)\\

\hline
\end{tabular}
\end{center}

\begin{center}
 {\bf Table 9.} Ordinal compatibility\\
\begin{tabular}{| l | c   c   c  c  c  c  |}
\hline
 &
 \(\widehat{D}_{1}\) & \(\widehat{D}_{2}\) & \(\widehat{D}_{3}\) &
 \(\overline{D}_{1}\) & \(\overline{D}_{2}\) & \(\overline{D}_{3}\) \\
\hline

 \(\widetilde{D}_{1}\)  &
 \(1\) & \(1\) & \(1\) &
 \(1\) & \(1\) & \(1\) \\

 \(\widetilde{D}_{2}\) &
 \(1\) & \(1\) & \(3\) &
 \(1\) & \(1\) & \(3\) \\

 \(\widetilde{D}_{3}\) &
 \(1\) & \(1\) & \(2\) &
 \(1\) & \(1\) & \(2\) \\

  \(\widehat{D}_{1}\) &
 & & &
 \(1\) & \(1\) & \(1\) \\

  \(\widehat{D}_{2}\) &
 & & &
 \(1\)  & \(1\) & \(1\) \\

  \(\widehat{D}_{3}\) &
 & & &
 \(1\)  & \(1\) & \(3\) \\

\hline
\end{tabular}
\end{center}

\section{Conclusion}

 The paper describes our first integrated step to combinatorial evolution and forecasting
 of requirements to communications.
 A hierarchical morphological model of the requirements system is
 used. Forecasting is based on
 morphological design approach
 (selection/composition of the best change items
 while taking into account their compatibility).
 The future research directions can involve the following:
  examination of the suggested approach
 to generations of wireless communications systems and the requirements to the systems.

\section{Acknowledgments}

%
 The work is partially supported by
 grant of
  Russian Science Foundation
 (project of Inst. for Information Transmission Problems
  14-50-00150
   ``Digital technologies and their applications'').


\end{document}